\documentclass[twocolumn,showpacs,jcp,superscriptaddress]{revtex4}
\usepackage{graphicx}
\usepackage[usenames]{color}
\usepackage{amssymb}
\usepackage{amsmath}
\usepackage{amsfonts}
\usepackage{mathrsfs}

\renewcommand{\epsilon}{\varepsilon}
\newcommand{\figurewidth}{0.45\textwidth}
\newcommand{\narrowfigurewidth}{0.40\textwidth}
\begin{document}
\title{Translocation of stiff polymers through a nanopore driven by binding particles}

\author{Wancheng Yu}
%\altaffiliation[]{Author to whom the correspondence should be addressed}
%\email{ywcheng@mail.ustc.edu.cn}
\affiliation{CAS Key Laboratory of Soft Matter Chemistry, Department of Polymer
Science and Engineering, University of Science and Technology of China, Hefei, Anhui
Province 230026, P. R. China}

\author{Yiding Ma}
%\altaffiliation[]{Author to whom the correspondence should be addressed}
%\email{ywcheng@mail.ustc.edu.cn}
\affiliation{CAS Key Laboratory of Soft Matter Chemistry, Department of Polymer
Science and Engineering, University of Science and Technology of China, Hefei, Anhui
Province 230026, P. R. China}

\author{Kaifu Luo}
\altaffiliation[]{Author to whom the correspondence should be addressed}
\email{kluo@ustc.edu.cn}
\affiliation{CAS Key Laboratory of Soft Matter Chemistry, Department of Polymer
Science and Engineering, University of Science and Technology of China, Hefei, Anhui
Province 230026, P. R. China}

\date{\today}
%%%%%%%%%%%%%%%%%%%%%%%%%%%%%%%%%%%%%%%%%%%%%%%%%%%%%%%%%%%%%%

\begin{abstract}

We investigate the translocation of stiff polymers in the presence
of binding particles through a nanopore by two-dimensional Langevin dynamics simulations.
We find that the mean translocation time shows a minimum as a function
of the binding energy $\epsilon$ and the particle concentration $\phi$, due to the interplay
of the force from binding and the frictional force.
Particularly, for the strong binding the translocation proceeds with
a decreasing translocation velocity induced by a significant
increase of the frictional force.
In addition, both $\epsilon$ and $\phi$ have an notable
impact on the distribution of the translocation time.
With increasing $\epsilon$ and $\phi$, it undergoes a transition from
an asymmetric and broad distribution under the weak binding to a
nearly Gaussian one under the strong binding, and its width becomes
gradually narrower.

\end{abstract}

%\pacs{87.15.A-, 87.15.H-}
%%%%%%%%%%%%%%%%%%%%%%%%%%%%%%%%%%%%%%%%%%%%%%%%%%%%%%%%%%%%%%

\maketitle
\section{Introduction}

The translocation of proteins and nucleic acids through a nanopore
is of essential importance to life, representative examples
including the passage of messenger RNA through nuclear pores,
post-translational transport of proteins across the endoplasmic
reticulum membrane, and virus injection \cite{Alberts}. In a seminal
experiment, Kasianowicz \textit{et al.} \cite{Kasianowicz} have
shown that single-stranded DNA and RNA can transverse the
water-filled $\alpha$-hemolysin channel, which is signaled by a
blockade in the channel ionic current. Since then, polymer
translocation through nanopores has garnered high-profile attention
for its far-reaching technological potential, such as rapid DNA
sequencing, gene therapy and controlled drug delivery
\cite{Kasianowicz,Branton,Turner,Gerland}.

In addition to its biological and technological relevance, polymer
translocation is also an important issue purely from the view of
polymer physics. Compared to an unconstrained case, the passage of a
polymer through a nanopore greatly reduces its degrees of freedom
and thus requires a force to overcome the energy barrier.
One of the main forms of such driving forces both \textit{in vivo} and
\textit{in vitro} is the trans-membrane force. In biological cells,
it comes from the trans-membrane electrical potential, while this is
achieved by an electric field mainly falling off the pore in
experiments, making use of the simple fact that biopolymers, such as
ssDNA, are negatively charged. This particular type of translocation
has been investigated extensively through experiments
\cite{Kasianowicz,Li,Storm,Wanunu} and theoretical works
\cite{Muthukumar,Sung,Lubensky,Sebastian,Klafter,Ambj,Kantor,Luo2,Hatlo}.
Another mechanism utilizes a chemical potential gradient
across the membrane. Examples include translocation of chains under different solvent conditions \cite{Liao,Weiss,Bhattacharya},
and that under the chemical potential gradient due to the binding particles (BPs) on the two sides of the
membrane \cite{Sung,Matlack,Matlack1,Brunner,Simon,Salman,Farkas,Schneider,Peskin,Glick,Zandi,Paolo,Elston,Liebermeister,Metzler1,Metzler2,Metzler3,Abdolvahab1,
Abdolvahab2,Abdolvahab3,Orsogna,Krapivsky,Yu}.
In what follows, we concentrate on polymer translocation in the presence of
BPs. Particularly, the chemical potential gradient in this work is induced by the
binding particles which only exists on the $trans$ side of the
membrane, as shown in Fig. \ref{Fig1}.

As to the translocation mechanism in the presence of BPs, two model have been
proposed. Simon \textit{et al.} \cite{Simon}
suggested that the translocation of chains is a simple thermal ratchet process, i.e.,
the role of BPs bound to the translocating chain is only to prevent it from moving backward, called the Brownian ratcheting mechanism.
Later, based on the results from Brownian molecular dynamics simulations, Zandi \textit{et al.} \cite{Zandi} considered that the binding of
BPs onto the chain can provide a mechanical force capable of pulling the chain through the pore, namely, the translocation is a
force-driven process.
Additionally, some important aspects associated with this problem have been
investigated successively, including the ``parking lot effect" due
to the size difference between BPs and chain monomers
\cite{Metzler1,Metzler2,Metzler3,Abdolvahab1,Abdolvahab2,Abdolvahab3},
the sequence dependence \cite{Abdolvahab1,Abdolvahab2,Abdolvahab3},
and the chain flexibility \cite{Yu}.

Although the studies mentioned above have provided a plenty of
creative insights into the translocation driven by BPs, the
underlying translocation dynamics even for a stiff polymer still
remains unclear. In previous simulations by Zandi \textit{et al.} \cite{Zandi}, the chain length
used is fixed at $N=16$, which is too short, and the binding energy
is fixed at $5 k_BT$.
Therefore, it is very necessary to understand the
influence of the binding energy $\epsilon$, the particle
concentration $\phi$, as well as the chain length $N$ on the
translocation dynamics.

To this end, we investigate the dynamics of a stiff polymer
translocation through a nanopore in the presence of BPs using
Langevin dynamics simulations. In section II,  We briefly elaborate
our model and the simulation technique. In section III, we present
our results and corresponding discussions. Finally, we give a
summary in section IV.

\section{Model and methods} \label{chap-model}%%%%%%%%%%%%%%%%%%%%%%%%%

\begin{figure}
\includegraphics*[width=\narrowfigurewidth]{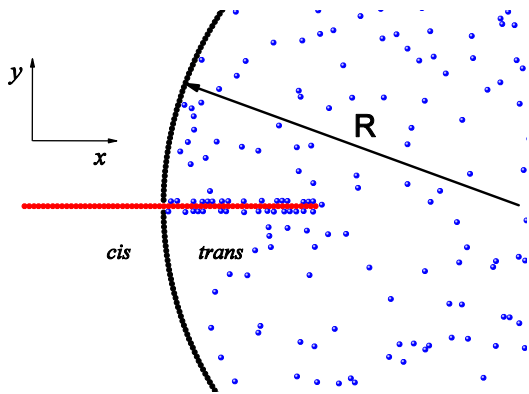}
\caption{(Color online) Schematic view of a stiff chain
translocating through a nanopore in presence of binding particles depicted
by blue particles inside the cell. The width of the pore is
$w=1.6\sigma$. The radius of the simulative cell is $R=73.5\sigma$.
        }
\label{Fig1}
\end{figure}

The model geometry we consider throughout this work is illustrated
in Fig. \ref{Fig1}, where a stiff polymer chain, modeled as a
bead-spring chain of Lennard-Jones (LJ) particles with the Finite
Extension Nonlinear Elastic (FENE) potential, is translocating
through a nanopore in the presence of BPs. Covalently bonded
monomers between nearest neighbor beads along the chain interact via
the FENE potential:
\begin{equation}
U_{FENE}(r)=-\frac{1}{2}kR_0^2\ln(1-r^2/R_0^2),
\label{eq1}
\end{equation}
where $r$ is the distance between consecutive monomers, $k$ is the
spring constant and $R_0$ is the maximum allowed separation between
connected monomers. We employ a short-range repulsive LJ potential
to incorporate the excluded volume interactions between chain
segments:
\begin{equation}
U_{LJ}(r) = \left\{ \begin{array}{ll}
4\epsilon_0[{(\frac{\sigma}{r})}^{12}-{(\frac{\sigma}{r})}^6] +\epsilon_0, & \textrm{$r\leq 2^{1/6}\sigma$}\\
0, & \textrm{$r>2^{1/6}\sigma$}.\\
\end{array} \right.
\label{eq2}
\end{equation}
Here, $\sigma$ is the diameter of a bead, $\epsilon_0$ is the depth of the potential, and $r$ is the distance between monomers.

Mobile BPs with the same size $\sigma$ as chain segments are
modeled as particles freely diffusing within a circular cell of the
radius $R=73.5\sigma$ with a pore of width $w=1.6\sigma$, and they
repel each other during the simulation via the same short-range
repulsive LJ potential (Eq. 2). The circular cell, consisting of one
monolayer stationary LJ particles, is repulsive for BPs to
prevent their overflow. The affinity between chain monomers and
BPs is achieved by exerting an attractive LJ potential with a
cutoff $2.5\sigma$ and the binding energy $\epsilon$. The particle
concentration $\phi$ is defined as the particle number density in
the cell, $\phi=N_{bp}/({\pi R^2})$. Note that, in our simulations,
the chain is treated as a completely straight one so that we neglect
the bending of the chain due to the binding of particles. In
addition, the pore is considered to be completely inert, namely,
having no impact on the chain other than to permit its
one-dimensional fluctuations in and out of the cell.

In the Langevin dynamics simulations, each mobile particle is subjected to conservative, frictional and random forces, respectively:
\begin{equation}
m{\bf \ddot{r}}_i=-{\bf\nabla}U_i-\xi {\bf v}_i+{\bf F}_i^R,
\label{eq3}
\end{equation}
Here, $U_i=\sum_{i\neq j}{U_{LJ}^{ij}}+U_{FENE}(i-1,i,i+1)$ for
chain monomers and $U_i=\sum_{i\neq j}{U_{LJ}^{ij}}$ for BPs,
$m$ is the particle's mass, which is assumed to be the same for the
monomer and the binding particle, $\xi$ is the particle's frictional
coefficient, ${\bf v}_i$ is the particle's velocity, and ${\bf
F}_i^R$ is the random force which satisfies the
fluctuation-dissipation theorem \cite{Chandler}. The system energy,
length and mass scales are determined by the LJ parameters
$\epsilon_0$, $\sigma$ and  bead mass $m$, leading to the
corresponding time scale $t_{LJ}=(m\sigma^2/\epsilon_0)^{1/2}$ and
force scale $\epsilon_0/\sigma$, which are order of ps and pN,
respectively. The reduced parameters for all simulations in the
present work are chosen to be $R_0=1.5$, $k=30$, $\xi=0.7$ and
$T=1.2$. Those chosen parameters give rise to an effective bond
length $\overline{\ell}$ of around 0.96$\sigma$.

The Langevin equation is integrated in time by the method proposed by Ermak and
Buckholz \cite{Ermak} in one dimension for the chain, and in two dimensions for
the binding particles. This integral scheme was also used by Zandi \textit{et al.} \cite{Zandi}.
In this way, fluctuations of chain monomers along the $y$ axis are forbidden
so that the polymer can only move along the $x$ axis, and such a polymer chain is
considered to be a rod-like chain.
This integral scheme means that we project any force on the rod along the $x$-axis and just
neglect the $y$-component. The approximation of neglecting $y$-components of the
forces does not change the results. Moreover, if the pore is long enough, the transverse
motion of the rod is not allowed.

Initially, the first monomer is placed just at the enter of the pore
($x=0$) and is kept fixed, and the rest of the chain and BPs
are under thermal collisions described by the Langevin thermostat to
reach the equilibrium state of the system. Then, the first monomer
is released and both the forward and backward movements of the chain are possible. If the first monomer
returns to the \emph{cis} side, it would be regarded as an
unsuccessful trial. A trial is considered to successful as the last
monomer exits the pore ($x=0.5$ for the first time) and such a trial
is called translocation. The translocation time $\tau$ is defined as
the time interval between the release of the first monomer and the
exit of the last monomer. Typically, we perform 1000 successful runs
to get an ensemble average in order to reduce statistical errors.

\section{Results and discussions} \label{chap-results}%%%%%%%%%%%%%%%%%%

In our simulations, we find that the translocation probability $P_{tran}$, which is defined as the
fraction of runs leading to successful translocation at given conditions, increases with
increasing the binding energy $\epsilon$ and the particle concentration $\phi$.
For example, at a given $\phi=0.8\%$, we observe $P_{tran}$=0.02, 0.20 and 0.44 for $\epsilon=1.5$, 4.5 and 7.5, respectively.
In addition, at a fixed $\epsilon=1.5$, $P_{tran}$ increases from 0.02 for $\phi=0.8\%$ to 0.14 for $\phi=19.2\%$.

\subsection{Mean first passage time $\tau(s)$ as a function of the translocation coordinate $\emph{s}$}

\begin{figure}
\includegraphics*[width=\figurewidth]{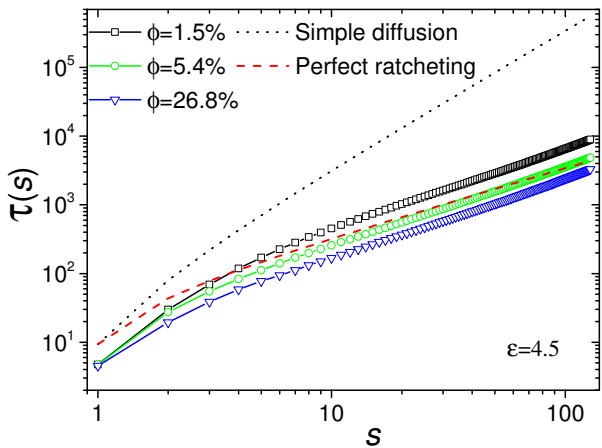}
\caption{(Color online) Mean first passage time $\tau(s)$ as a function of the translocation coordinate $\emph{s}$. The chain length is $N=128$.
Mean first passage times for the simple diffusion and the perfect ratcheting motion are plotted as the dotted line (black) and the short dashed line (red), respectively.
        }
\label{Fig2}
\end{figure}

We define $\tau(s)$ as the mean first passage time for $\emph{s}$
monomers exiting the pore. Fig. \ref{Fig2} show $\tau(s)$ for chain
length $N=128$, binding energy $\epsilon=4.5$ and different
particle concentrations $\phi$. To make a comparison, we also
calculate the times needed for $\emph{s}$ monomers exiting the pore
by the simple diffusion and the perfect ratcheting motion, denoted
as $\tau_{diff}(s)$ and $\tau_{ratchet}(s)$, respectively. Note
that, due to the initialization condition, $\tau(s=1)$ is the time
for the first monomer to first arrive the pore exit. Thus,
$\tau_{diff}(s)=\frac{[0.5+(s-1)\overline{\ell}]^2}{2D}$ (the black
dotted line), and
$\tau_{ratchet}(s)=\frac{0.5^2+(s-1)\overline{\ell}^2}{2D}$ (the
red dashed line), with $D=\frac{k_BT}{N\xi}$ being the diffusion
coefficient of the whole chain.

As expected, $\tau_{diff}(s)$ is significantly longer than the mean
first passage time $\tau(s)$ for different particle concentrations.
At a low particle concentration ($\phi=1.5\%$), the translocation is slower than the perfect
ratcheting motion.
However, with increasing $\phi$ from $1.5\%$ to $26.8\%$, the
whole curve lies under $\tau_{ratchet}(s)$, demonstrating that the translocation is faster than the perfect
ratcheting motion. This behavior is a little different from the theoretical
prediction where the perfect ratcheting result is considered as the lower limit to
the translocation time \cite{Metzler2}. The reason may be from the
non-equilibrium effect of the translocation process. We should point out
that our results are in agreement with the numerical findings by
Zandi \textit{et al.} \cite{Zandi}, where the translocation in the
presence of BPs is further attributed to a force-driven process.

Interestingly, at a moderate $\phi=5.4\%$, the translocation is
faster than the the perfect ratcheting motion at first, but the
opposite is the case after $s \geq 85$. This dynamical behavior
indicates a striking chain length dependence in this issue.
%when the effective chaperone concentration reduces markedly during the translocation.
Zandi \textit{et al.} \cite{Zandi} have not observed
this phenomenon possibly due to the too short chain used in their simulations
($N=16$).

\subsection{Influence of $\epsilon$ and $\phi$ on the translocation time}

\begin{figure}
\includegraphics*[width=\figurewidth]{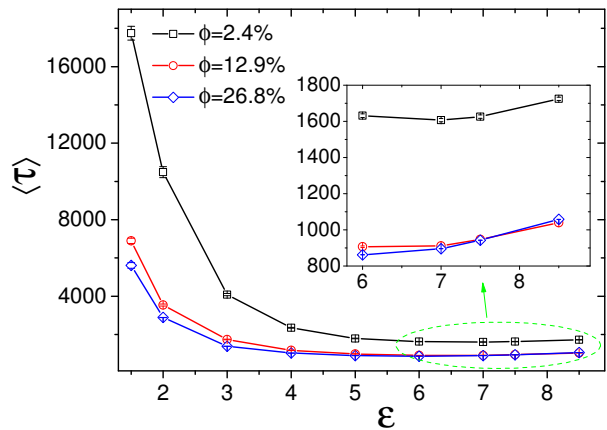}
\caption{(Color online) The mean translocation time as a function of
the binding energy for three different particle concentrations
$\phi$ =2.4\%, 12.9\%, and 26.8\%. The inset is a zoom of the data
for large $\epsilon$. The chain length here is $N=64$. The solid
lines are guides to the eye, and almost all the errors of the data
are smaller than the sizes of symbols in the plot.
        }
\label{Fig3}
\end{figure}

\begin{figure}
\includegraphics*[width=\figurewidth]{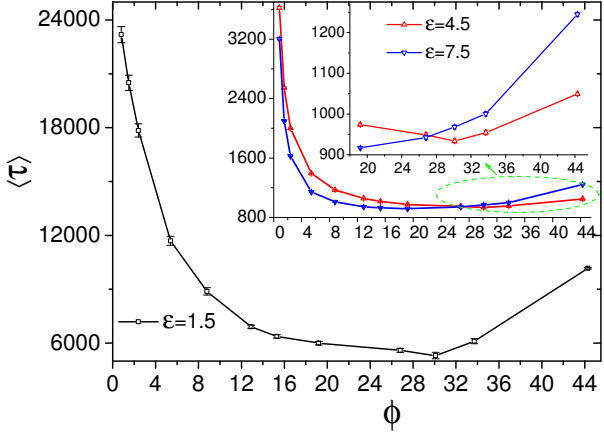}
\caption{(Color online) The mean translocation time as a function of
the particle concentration for three different  binding energies
$\epsilon$ =1.5, 4.5, and 7.5. The chain length here is $N=64$. The
solid lines are guides to the eye, and almost all the errors of the
data are smaller than the sizes of symbols in the plot.
        }
\label{Fig4}
\end{figure}

Next, we examine the influence of $\epsilon$ and $\phi$ on the
translocation dynamics by measuring the mean translocation time
$\langle\tau\rangle$. Fig. \ref{Fig3} shows that for low, moderate
and high $\phi$, $\langle\tau\rangle$ decreases rapidly at first,
and then slowly approaches a minimum with increasing $\epsilon$.
Afterwards, a slight but non-negligible increase is observed, see
the inset of Fig. \ref{Fig3}. Furthermore, the decay rate gradually
diminishes before reaching the minimum.
A similar effect of $\phi$ on $\langle\tau\rangle$ is observed as
shown in Fig. \ref{Fig4}.

To understand these behaviors, we first consider the relevant
timescales and the force from binding.  There are three relevant
timescales: the time $\tau_{0}=N\delta^2/2D_0$ for a chain to
diffuse a distance of a binding site length $\delta$, with
$D_0=k_BT/\xi$ being the diffusion coefficient of a monomer; the
typical time $\tau_{unocc}$ that a binding site keeps vacant, and
the characteristic time $\tau_{occ}$ that a binding site stays
occupied. The typical time $\tau_{unocc} \sim R_{bp}^2/D_{bp} = 1/(\phi
D_{bp})$, where $R_{bp}\sim \phi^{-1/2}$ is the average distance
between BPs in solution in two dimensions and $D_{bp}$ is the
diffusion coefficient of BPs. Taking into account the same
size for a binding particle and a monomer, we have $D_0=D_{bp}$. The
relationship between $\tau_{unocc}$ and $\tau_{occ}$ meets
$\tau_{occ} = \kappa\tau_{unocc} \sim \frac{A_0}{1-\phi
A_0}\exp(\epsilon/k_BT)/D_{bp}$ \cite{Metzler1}. Here,
$\kappa=\frac{\phi}{1-\phi A_0} K_{eq}$ is a relevant measure of
the effective binding strength, $K_{eq}=A_0\exp(\epsilon/k_BT)$ is
the equilibrium binding constant with $A_0=\pi\sigma^2/4$ being the
typical binding particle area. Then, three dynamical regimes were divided
according to three relevant timescales \cite{Metzler1,Metzler2}: the
diffusive regime for slow binding ($\tau_{0}\ll \tau_{occ},
\tau_{unocc}$), the reversible binding regime for fast binding and
unbinding ($\tau_{0}\gg \tau_{occ}, \tau_{unocc}$), and the
irreversible binding regime for fast binding but slow unbinding
($\tau_{unocc}\ll \tau_{0}\ll \tau_{occ}$).

For the reversible binding, by calculating the binding partition
function,  Ambj\"{o}rnsson \textit{et al.} \cite{Metzler1,Metzler2}
have obtained the force $F_B$ (in units of $k_BT/\sigma$) exerted on
the chain by BPs. Considering the univalent binding and the
cooperativity effects between BPs bound to the chain, for
large translocation coordinate $s$, the finite size corrections
become irrelevant and $F_B$ remains a constant value. $F_B$ is given
as
\begin{equation}
F_B(\kappa)\approx\ln\{\frac{1+\omega\kappa}{2}+[(\frac{1+\omega\kappa}{2})^2+(\omega-1)\kappa]^{1/2}\},
\label{eq4}
\end{equation}
Here, $\omega$ is a cooperativity parameter. In the present work,
since BPs bound to the chain interact repulsively, $0< \omega
< 1$ if $\kappa \leq 1/4$, and $1/2 \leq \omega < 1$ for all
$\kappa$ values \cite{Metzler1}. Evidently, $\kappa$ increases with
$\phi$ and $\epsilon$, and thus Eq. (\ref {eq4}) indicates that
$F_B$ increases with the binding energy $\epsilon$ and the particle
concentration $\phi$.
This is the reason that $P_{tran}$ increases with increasing $\epsilon$ and $\phi$.
For irreversible binding, however, BPs do not have time to
unbind during the translocation, which leads to the thermodynamic
evaluation of the force inapplicable.

Our results shown in Figs. \ref{Fig3} and \ref{Fig4} can be
qualitatively explained from the perspective of the forces acting on
the chain during the translocation, including the force from binding
$F_B$ and the frictional force $F_{fric}$. The initially sharp
descent of $\langle\tau\rangle$ with increasing $\epsilon$ and
$\phi$ is dominated by a substantial increase in $F_B$. However,
the increases in $\epsilon$ and $\phi$ also simultaneously lead to
a larger $\tau_{occ}$.
As a result, $F_{fric}$ rises because segments with bound BPs have
a higher friction coefficient than free segments \cite{Simon}, which
partially counteracts the favorable factor for the translocation
caused by the increasing $F_B$. This fact is also the origin of the
gradual reduction in the decay rate.

\begin{figure}
\includegraphics*[width=\figurewidth]{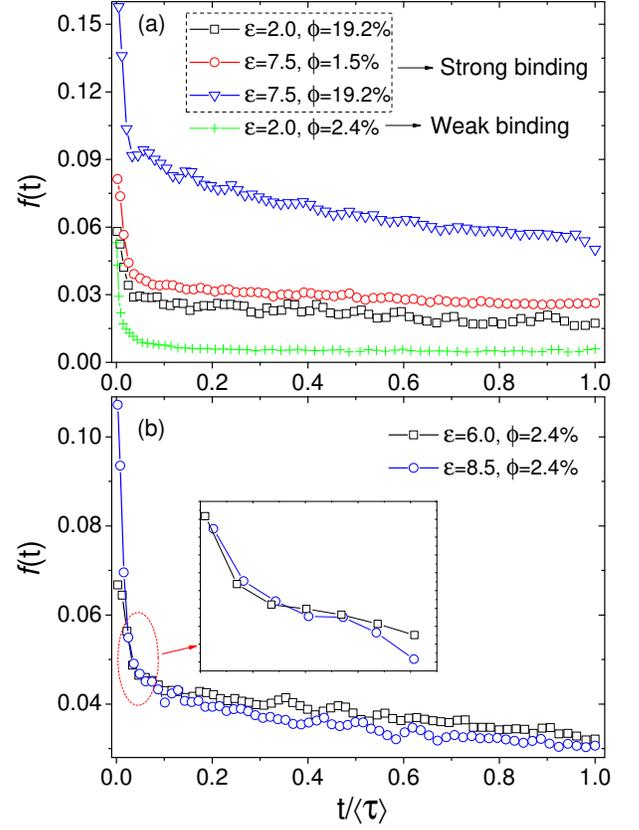}
\caption{(Color online) Evolution of the polymer flux through the
nanopore $f(t)$ with the time normalized by the mean value
for (a) the weak and strong binding, and (b) two different binding
strengths in the strong binding regime. The inset is a zoom of the
data in the intersection area of the two curves. The solid lines are
guides to the eye. The chain length here is $N=64$.
        }
\label{Fig5}
\end{figure}

To further understand why $\langle\tau\rangle$ has a minimum as a
function of $\epsilon$ and $\phi$, we have calculated the polymer
flux through the nanopore $f(t)$ during the translocation
Here, $f(t)=\frac{d\langle s(t)\rangle}{dt}$,
defined as the differentiation of the average translocation
coordinate $\langle s(t)\rangle$ with respect to the time,
characterizes the instantaneous speed of the translocation process,
namely, the number of segments passing through the pore per unit
time.

When the binding is weak, it is reasonable to deem that the effect
of BPs' binding on the friction coefficient of the chain is
negligible because the unbinding of bound BPs is quite fast.
Therefore, the translocation under this case would be in steady state.
However, as the binding gets too strong, $F_{fric}$ becomes the
prominent factor, leading to a gradually decreasing translocation
velocity, as seen in Fig. \ref{Fig5}(a).

With the two competing effects caused by increasing $\epsilon$ and
$\phi$ operative, we speculate that too strong binding is in
favor of the translocation in its primary stage, and it then results
in a more significant increase in $F_{fric}$ and the ensuing more
dramatic decrease in the translocation velocity. Fig. \ref{Fig5}(b)
shows that at a given $\phi=2.4\%$, the polymer flux of the
translocation under $\epsilon=8.5$ is indeed larger than that of the
weaker one ($\epsilon=6.0$) at the beginning, but with a faster
decay rate. Finally, about $95\%$ of the translocation under the
stronger binding proceeds with a slower velocity, and consequently a
longer translocation time is required.

Zandi \textit{et al.} \cite{Zandi} showed that the mean translocation time decreases monotonically with the
particle concentration increasing from $1.1\%$ to $44.2\%$.
In contrast, our results indicate that the mean translocation time has a minimum as a
function of $\phi$ which increases from $0.8\%$ to $44.2\%$. The
difference is due to the too short chain length used in their simulations. As a result, the translocation
has not entered into the frictional force-dominant regime.

\subsection{Distribution of the translocation time}

\begin{figure}
\includegraphics*[width=\figurewidth]{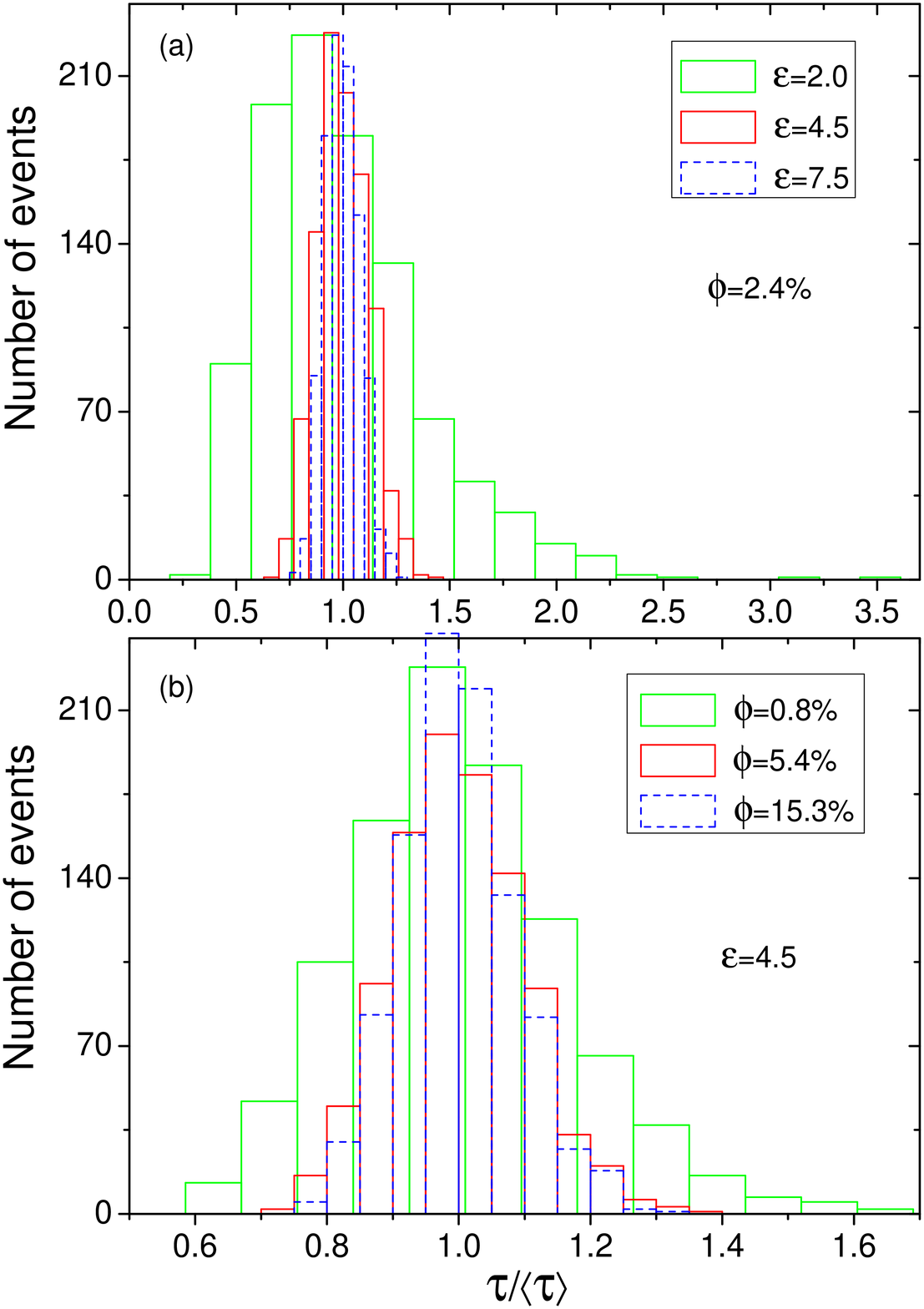}
\caption{(Color online) The influence of (a) the binding energy
$\epsilon$ and (b) the particle concentration $\phi$ on the
distribution of 1000 translocation time for a chain of length $N=64$
normalized by their mean value $\langle\tau\rangle$.
        }
\label{Fig6}
\end{figure}

Previous studies have shown that translocation driven by a transmembrane electric
field \cite{Kantor,Luo3,Luo4} or a constant pulling force exerted
on the first monomer \cite{Luo5} has a qualitatively different
shape of the distribution of translocation time compared to that of
the unbiased translocation case \cite{Luo4,Chuang}. Now that the
polymer translocation in the presence of BPs is a force-driven process
and the driving force $F_B$ is a function of the binding energy
$\epsilon$ and the particle concentration $\phi$, it is natural
to wonder whether changes in $\epsilon$ and $\phi$ will have
similar effects on the distribution of the translocation time.

As sketched in Fig. \ref{Fig6}(a), the distribution for
$\epsilon=2.0$ is asymmetric and broader, while the histogram obeys
nearly Gaussian distribution as $\epsilon$ increases. Fig.
\ref{Fig6}(b) shows that at a relatively large binding energy
$\epsilon=4.5$, almost all of the histograms approach nearly
Gaussian distributions, even under a fairly low $\phi=0.8\%$. And
the width of the distribution becomes narrower as the binding gets
stronger. These results can be ascribed to the simple fact that
$F_B$ increases with increasing $\epsilon$ and $\phi$, and are
qualitatively similar to the observations reported by previous
studies \cite{Luo4,Kantor,Luo5}. What's more, the distribution of
the translocation time is more sensitive to $\epsilon$ than $\phi$
in that $\epsilon$ has a more significant effect on the effective
binding strength $\kappa$. At a smaller $\epsilon=2.0$, we could
still observe the transition from an asymmetric distribution to a
nearly Gaussian one as $\phi$ increases.

Most recently, it has been demonstrated that the probability density
function (PDF) of the translocation time is solely determined by the
dimensionless P\'{e}clet number, denoted as $Pe$, \cite{Abdolvahab3}.
It is a dimensionless parameter comparing drift strength and diffusivity.
The PDF changes from a broad distribution in the diffusion dominated regime at small
$Pe$ to a significantly narrower distribution in the regime of drift
domination at large $Pe$. We calculate the effective P\'{e}clet
number, $Pe(\phi_c, \epsilon)$ as follows,
$Pe(2.4\%,2.0)\approx8.66$, $Pe(2.4\%,4.5)\approx15.30$,
$Pe(2.4\%,7.5)\approx23.38$, $Pe(0.8\%,4.5)\approx11.55$,
$Pe(5.4\%,4.5)\approx18.18$, and $Pe(15.3\%,4.5)\approx 21.83$.
Obviously, $Pe(2.4\%,2.0)\approx8.66$ for the broader distribution
is smaller than these for significantly narrower distributions.
Therefore, our results about the distribution of the translocation
time are qualitatively in agreement with the findings in Ref.
\onlinecite{Abdolvahab3}.

%%%%%%%%%%%%%%%%%%%%%%%%%%%%%%%%%
\section{Conclusions} \label{chap-conclusions}

In this work, we have investigated the translocation of stiff
polymers through a nanopore in the presence of BPs by
performing 2D Langevin dynamics simulations. We show that under a
certain $\epsilon$ and $\phi$, the translocation is faster than
the perfect ratcheting motion at first, but the opposite is the case
for the late stage of the translocation process. This indicates the
striking chain length dependence of the translocation dynamics.
By scanning a large range of the parameter spaces, we find that there exist an
optimal $\epsilon$ and an optimal $\phi$ for the translocation. Then, a
qualitative explanation is given from the perspective of the forces
acting on the chain. Finally, both $\epsilon$ and $\phi$ have an
notable impact on the distribution of the translocation time. Specifically,
with increasing $\epsilon$ and $\phi$, it undergoes a transition from
an asymmetric and broad distribution under the weak binding to a
nearly Gaussian one under the strong binding, and its width becomes
gradually narrower.

%%%%%%%%%%%%%%%
\begin{acknowledgments}
This work is supported by the National Natural Science Foundation of China (Grant Nos.
21225421, 21074126, 21174140, J1030412) and the `` Hundred Talents Program '' of CAS.
\end{acknowledgments}

%%%%%%%%%%%%%%%

\end{document}